# A Unique Theory of Gravity and Matter

George Chapline

Lawrence Livermore National Laboratory
Livermore, Ca 94550
chapline1@llnl. gov

## Abstract

The author has previously suggested that the ground state for 4-dimensional quantum gravity can be represented as a condensation of non-linear gravitons connected by Dirac strings. In this note we suggest that the low-lying excitations of this state can be described by a quasi-topological action of the form $\int d^{13}z$ $F_4 \wedge F_5 \wedge F_4$ , corresponding to a trilinear coupling of solotonic 8-branes and 7-branes. It is shown that when the excitations associated with $F_5$ are neglected, the effective action can be interpreted as a theory of conformal gravity in four dimensions. This in turn suggests that ordinary gravity as well supersymmetric matter and phenomenological gauge symmetries arise from the spontaneous breaking of topological invariance. The possibly deep mathematical significance of this theory is also noted.

# 1. Introduction

Although the discovery of superstring models led to great expectations that a theory unifying general relativity and elementary particle physics was in hand, it is by now clear that superstring models can at best be only an approximation to some more fundamental theory. One problem with superstring models is that they seem to require for their formulation a classical background spacetime whose very existence is inconsistent with a quantum theory of gravity. This suggests [1] that direct attempts to quantize gravity might be more fruitfull approach to a fundamental theory than superstring models. However because it has been generally felt that despite its shortcomings superstring theory itself offers the best hope for establishing a finite quantum theory of gravity little attention has been paid to this idea. On the other hand a finite quantum model for the ground state of 4-dimensional quantum gravity based on a condensate wavefunction for non-linear gravitons [2] provides an attractive explanation for how a classical spacetime background could emerge from the spacetime "foam" of quantum gravity. Furthermore geometric quantization [3] of the self-dual and anti-self-dual non-linear gravitons in this model induces degrees of freedom that resemble the light-cone excitations of a relativistic membrane. In fact recently it has been formally shown [4] that a truncation of the quantized self-dual gravity theory of ref.3 can be reinterpreted as a quantum theory of self-dual membranes in the light-cone gauge. Athough the exact nature of the degrees of freedom in a fundamental theory underlying superstring models is still rather obscure, there are strong hints [5] that the fundamental degrees of freedom are related in some way to those of a relativistic membrane. Thus we are led to suspect that the condensate model of ref.2 can perhaps be regarded as not just a model for the ground state of quantum gravity but as a stepping stone to a fundamental theory underlying superstring models.

Our proposal in the following for a step towards this fundamental theory is motivated in part by the observation that the classical degrees of freedom of a membrane moving in D dimensions can be interpreted as the gauge potentials in a (D-1)-dimensional SU($\infty$) Yang-Mills theory that has been dimensionally reduced to one temporal dimension; i.e. SU($\infty$) quantum mechanics [6]. It has been shown [7,8] that in the case of self-dual (or anti-self-dual) membranes this SU($\infty$) quantum mechanics contains the 3-dimensional Toda model that was used in ref.2 to represent classical self-dual (or anti-self-dual) gravitons. Coincidently this same 3-dimensional Toda model had been previously obtained as a byproduct of a scheme [9] to generalize an exactly solvable model of anyon superconductivity to three dimensions by replacing the ordinary Maxwell gauge potential in the 2-dimensional anyon model with an SU(∞) gauge potential. The spatial positions and gauge parameters of the SU(∞) anyons or "chirons" in this model can be reinterpreted [9,10] as the moduli of the self-dual or anti-self-dual Einstein spaces, as well as the moduli of classical solutions to the SU(∞) Nahm equations (cf. ref.7). Thus it would appear that the condensate vacuum model of ref.2 serindipitously incorporates degrees of freedom very much like what is thought to be needed in a theory underlying superstring models. It is noteworthy in this respect that the condensate model of ref.2 goes beyond known classical solutions of the Einstein or Nahm equations because it encorporates both self-dual and anti-self-dual degrees of freedom, and furthermore these two types of degrees of freedom play roles much like the left-moving and right-moving sectors of superstring theory. On the other hand one seeming deficiency of the condensate model of ref.2 in comparison with superstring theories is that it does not encorporate in any obvious way matter-like degrees of freedom.

In order to progress beyond the simple model of ref.2 we would like first of all to take notice of the fact that the continuous Toda model representation for self-dual membranes has natural realizations in 7+1 dimensions [11,12]. Therefore we are led to the direct product of two 8-dimensional manifolds, or perhaps the complexification of an 8-manifold, as the natural setting for the paired non-linear gravitons appearing in the condensate model of ref.2. In addition to the two 8-dimensional manifolds that we are assuming are required for the description of the paired non-linear gravitons, the Dirac strings appearing in the condensate model of ref.2 naively require an extra dimension; however in keeping with the "F-theory" interpretation of superstring compactifications with varying dilaton and axion fields [13] and the related geometric interpretation of N=1 quantum Yang-Mills theories [14], in the following we will replace these Dirac strings with 7-branes. Counting up the number of dimensions required we are led to 24+2 dimensions as the natural number of spacetime dimensions in which to formulate an action principle for describing the low energy excitations of the condensate vacuum of ref.2. We will tentatively identify the 13-dimensional spaces that play a central role in many examples of F-theory compactifications [15] with real sections of our 26-dimensional space. Because the "U-manifolds" that occur in F-theory compactifications are typically themselves complex manifolds, it can be expected that quaternionic representations for the spacelike piece of the 26-dimensional space will also play some role.

An additional hint that 13-dimensions plays an important role in a fundamental theory is provided by the fact that certain low energy interactions of Type II superstrings compactified on a Calabi-Yau manifold 3-fold can be represented by a Chern-Simons-like effective lagrangian in 12-dimensions [16]. Just as the Chern-Simons form in 3-dimensions can often be regarded as the dimensional reduction of the Pontryagin form for 4-dimensional manifolds, so we might also assume that the effective action of ref.16 points is the reflection of a more fundamental action in 13 dimensions. Our main purpose in this paper is to draw attention to the apparently unique significance of this topological action in thirteen dimensions. Of course since we believe that the setting for our effective action should be a complexification of the 13-dimensional space of F-theory, our action will be only quasi-topological.

Motivated then by a desire to represent the low-lying excitations of the condensate vacuum of Ref.2 in a way consistent with F-theory we introduce the following effective action:

$$I = \int_{M_{13}} d^{13}z \; F_4 \wedge F_5 \wedge F_4 , \qquad (1)$$

where $F_4$ and $F_5$ are the field strengths for 3-form and 4-form gauge fields defined on a 13-dimensional complex manifold $M_{13}$. It is easy to see by integrating by parts and using the Bianchi identity that (1) is invariant under infinitesimal variations in either $F_4$ or $F_5$, so that the action (1) is invariant with respect to diffeomorphisms of real sections of $M_{13}$. In addition to topological gauge invariance the action (1) will be invariant under ordinary gauge transformations of the 3-form gauge field associated with $F_4$ and the 4-form gauge field associated with $F_5$. As mentioned above the action principle (1) is a natural generalization of a 12-dimensional effective action which describes the low energy couplings of certain vector and tensor multiplets in F-theory. However, in contrast with the discussion of ref.16 where it is not contemplated that ordinary gravity should be described by Chern-Simons-like interactions between 3-form and 4-form gauge fields, we will show in the following section that the tensor couplings implied by the action (1) include interactions that can be used to describe conformal gravity in 4-dimensions. We will argue in sections 3 and 4 that in addition matter and Yang-Mills gauge fields perhaps similar to those of the Standard Model will appear when the topological invariance of (1) is spontaneously broken.

## 2. An 8-dimensional formulation for general relativity

One of the motivating factors behind the construction of the condensate model for the vacuum state of quantum gravity was that it should have lowlying excitations resembling those of classical conformal gravity. We will now show that in the limit where excitations associated with the 5-form $F_5$ are neglected the action (1) can be reinterpreted as a theory of gravity in four dimensions. If we assume that $F_5$ is a delta function of the 5 coordinates corresponding to its 5 indices, then the action (1) becomes a quasi-topological guage theory in eight complex dimensions:

$$I_8 = \int d^8z \, F_4 \wedge F_4. \qquad (2)$$

The action $I_8$ is invariant under arbitrary coordinate transformations on real sections of the eight dimensional subspace of $M_{13}$ corresponding to $F_4 \wedge F_4$, as well as arbitrary variations in the 3-form gauge potentials $A_{\nu\rho\sigma}$ that vanish on the boundary of this real section.

In order to make contact with the condensate vacuum picture of ref. 2 we would like to somehow identify the 4-forms in (1) with self-dual and anti-self-dual graviton degrees of freedom. With this objective in mind one might be tempted to identify the 4-forms in (1) with the Weyl tensors for self-dual and anti-self-dual Einstein spaces. However such an interpretation would be inconsistent with our desire to identify real sections of $M_{13}$ with the 13-dimensional spaces used in F-theory. Instead we note that maintainence of an F-theory-like interpretation for our action on real sections of $M_{13}$ will require that the 4-forms in (1) encorporate contributions from both self-dual and anti-self-dual graviton (or alternatively self-dual and anti-self-dual membrane) degrees of freedom. As it happens Witten's twister interpretation of the classical Yang-Mills equations [17] together with a representation of the Weyl tensor as a "Yang-Mills" field strength for a twister connection [18] provide a natural formalism for including self-dual and anti-self-dual degrees of freedom in representations of curved 4-dimensional spacetime. In particular if the Weyl tensor $W_{\mu a \nu b}$ is self-dual with respect to indices $\mu, \nu$ on one 4-dimensional space, and anti-self-dual with respect to these coordinates on another 4-dimensional space, then it can be shown [19] that on the diagonal 4-dimensional space it will satisfy the equations for conformal gravity:

$$(\nabla^\mu \nabla^\nu - \frac{1}{2} R^{\mu\nu}) W_{\mu a \nu b} = 0, \qquad (3)$$

where $R^{\mu\nu}$ is the classical Ricci tensor.

As discussed in ref.s 11 and 12 there exist natural ways to extend an SO(4) gauge field $A_\mu^{ab}$ in 3+1 dimensions to a gauge field $A_\mu^{AB}$ in 7+1 dimensions if the SO(4) gauge group is replaced SO(8) gauge group. Furthermore there exist solutions to the SO(8) Yang-Mills equations in 7+1 dimensions such that when the 4-form $\mathrm{Tr} f_2 \wedge f_2$, where $f_2$ is the 2-form SO(8) field strength, is substituted into the integral (2), the result is proportional to the Euler number for the 8-dimensional manifold on which $f_2$ is defined [20]. This suggests that on a 7+1 dimensional subspace of $M_{13}$ we might choose as an ansatz for $F_4$

$$^{\pm}F_4 = \mathrm{Tr}\, ^{\pm}W \wedge ^{\pm}W, \qquad (4)$$

where $^{\pm}W$ is the field strength for the aforementioned extension to 7+1 dimensions of the twister connection for a self-dual (anti-self-dual) Einstein space. The $^{\pm}W$ apply on separate subspaces obtained by splitting the 7+1 complex dimensional subspace of $M_{13}$ into two pieces each with 7+1 real dimensions. We would naturally want to identify the section corresponding to F-theory with the diagonal subspace.

Our proposal that gravitational-like interactions in 4-dimensions from a topological action in 8-dimensions also sheds new light on the Ashtekar constraints in a Hamiltonian formulation of general relativity [21]. If we denote the electric field corresponding to the spatial component $A_i^a$ by $E_i^a$, then the generators of rotations of internal and spatial indices will be $\varepsilon_{ijk} A_j^a E_k^a$ and $\varepsilon_{abc} A_j^b E_k^c$ respectively. Topological invariance of the action implies invariance under these rotations, and it follows from identities given in ref.21 that imposing the constraint that physical states are singlets under these rotations is equivalent to the Ashtekar constraints provided $g^{ij} A_i^a E_j^a = 0$. Thus the Ashtekar constraints may be viewed as being a reflection of the topological invariance of (2), and this "explains" why these constraints do not require $g^{ij} \neq 0$.

In our picture of the excitations of the condensate vacuum it is the topological invariance of the effective action (2) that leads to general covariance with respect to coordinate transformations in four dimensions. On the other hand according to Einstein's theory of general relativity gravity is a reflection of the actual geometry of spacetime and not just its topology; therefore in order to explain ordinary gravity in four dimensions we must spontaneously break the topological invariance of the 8-dimensional action (2). Presumably one signal for this spontaneous breaking of topological invariance will be that the metric for 4-dimensional spacetime is non-vanishing; i.e. $g^{\mu\nu} \neq 0$, which as we have previously noted can be interpreted as the order parameter for the condensate vacuum. Whereas we gave a microscopic explanation for the appearance of this order parameter in ref.2, in the context of our effective action (1) the explanation for the appearance of this order parameter is equivalent to the ambitwister theory of gravity [20], where the familiar four dimensional spacetime of classical physics is interpreted as the space of null geodesics formed by the intersections of the self-dual and anti-self-dual null surfaces of non-linear gravitons.

### 3. Ordinary matter as topological ghosts ?

Even though the action principle introduced in eq. (1) was intended to describe the low lying excitations of our condensate model for the ground state for quantum gravity, we will now argue that it may also imply the existence of matter-like and gauge field-like degrees of freedom. We first show that topological gauge fixing of the action (1) leads to an interesting menu of ghost fields, whose fermionic components can perhaps be identified as matter fields. Regarded as a topological action in 13 dimensions (1) is invariant under the transformations

$$\delta A_{\mu\rho\sigma} = D_\mu \varepsilon_{\rho\sigma} + \varepsilon_{\mu\rho\sigma} \, , \, \delta A_{\mu\nu\rho\sigma} = D_\mu \varepsilon_{\nu\rho\sigma} + \varepsilon_{\mu\nu\rho\sigma} \, , \qquad (5)$$

where the first terms represent ordinary gauge transformations while $\varepsilon_{\mu\rho\sigma}$ and $\varepsilon_{\mu\nu\rho\sigma}$ are infinitesimal 3-forms and 4-forms that vanish on the boundary of 13-dimensional space. It should be noted that these parameters are only defined modulo derivative terms. In fixing the topological gauge of (1) we follow the lead of Balieau and Singer [22]. In particular we will assume as a "topological" gauge conditions that $F_5$ is the dual with respect to 13 dimensions of $F_4 \wedge F_4$:

$$F_5 = (F_4 \wedge F_4)^* \, . \qquad (6)$$

One can construct a BRST operator *s* corresponding to the combined symmetries (5) by introducing ghost fields associated with the gauge symmetries and gauge conditions as follows:

| $A_3$ | $A_4$ | $F_4$ | $F_5$ | $F_5 + {}^*(F_4 \wedge F_4)$ |
|---|---|---|---|---|
| $c_{\rho\sigma}$ $\bar{c}_{\rho\sigma}$ | $c_{\nu\rho\sigma}$ $\bar{c}_{\nu\rho\sigma}$ | $\psi_{\nu\rho\sigma}$ $\bar{\psi}_{\nu\rho\sigma}$ | $\psi_{\mu\nu\rho\sigma}$ $\bar{\psi}_{\mu\nu\rho\sigma}$ | $\bar{\chi}^{\mu\nu\rho\sigma\tau}$ |
| | | $\phi_{\rho\sigma}$ $\bar{\phi}_{\rho\sigma}$ | $\phi_{\nu\rho\sigma}$ $\bar{\phi}_{\nu\rho\sigma}$ | |

$$+ ...,  \qquad (7)$$

where the dots refer to additional ghost of ghost terms. In addition we will introduce the same set of ghost fields for the self-dual $F_4$'s as introduced in ref. 22 to gauge fix 4-dimensional topological Yang-Mills theory. For the gauge fixed lagrangian we propose

$$I = I_{top} + \int d^{13}z\, s\, \{\, \bar{\chi}^{\mu\nu\rho\sigma\tau} [F_5 + {}^*(F_4 \wedge F_4) + b_5]_{\mu\nu\rho\sigma\tau} + \bar{\chi}^{\mu\nu\rho\sigma}({}^*F_4 \pm F_4 + b_4)_{\mu\nu\rho\sigma}$$
$$+ \bar{c}^{\nu\rho\sigma}(\partial_\mu A_{\mu\nu\rho\sigma} + B_{\nu\rho\sigma}) + \bar{c}^{\rho\sigma}((\partial_\mu A_{\mu\rho\sigma} + B_{\rho\sigma}) \qquad (8)$$
$$+ ...$$

where $s$ is the BRST operator for the symmetries (5), the B's are auxillary fields associated with ordinary gauge invariance, while $b_4$ and $b_5$ are auxillary fields associated with topological deformations of $A_{\mu\nu\rho}$ and $A_{\mu\nu\rho\sigma}$. The dots in eq. (8) denote additional terms involving the ghost of ghost fields indicated in eq. (7). Elimination of the auxillary fields in the action (8) by guassian integration leads to a supersymmetric local field theory, with ordinary kinetic energy terms and where the BRST operator $s$ becomes a supersymmetry operator relating bosonic and fermionic fields.

Evidently then some of the ghost fields associated with the gauge invariances of our topological field theory can be interpreted as fermions. Except possibly in certain critical dimensions neither the bosonic nor fermionic ghost fields introduced in eq. (7) will decouple from the topological gauge fields $A_{\mu\nu\rho}$ and $A_{\mu\nu\rho\sigma}$, and it has been suggested [23] that both the bosonic and fermionic coordinates of superstrings arise in just this way. It is not a great leap to imagine that bosonic and fermionic coordinates for p-branes can arise in a similar way, and indeed it is clear from their role as lagrange multipliers for the classical constraint conditions which fix the topological gauge that the ghost fields associated with topological invariance are closely related to the zero modes for the 8-brane and 7-brane solitons associated with $F_4$ and $F_5$. As in the Green-Schwarz theory of superstrings we of course expect that the appearence of fermionic coordinates for the solotonic 8-branes and 7-brane appearing in the action (1) will be a signature for the appearence of ordinary space-time fermions.

## 4. Gauge symmetries as crystallographic symmetries

As is well known to specialists in group theory there is a close connection between the periodic arrangements of atoms in crystalline solids and the root diagrams for certain Lie algebras [24]. For example the root diagram for SU(4) can be identified with the unit cell for a face-centered cubic lattice. In this section we will argue that far from being just an isolated curiosity this connection actually underlies the appearance of phenomenological gauge symmetries. Our explanation for the appearance of characteristic gauge symmetries is actually closely parallel to the putative mechanism for the appearance of quantum Yang-Mills theories in F-theory. In particular quantum Yang-Mills theories arise from the geometric arrangement of parallel 7-branes wrapped around accumulations of vanishing 2-cycles [14]. Just why certain types of singularities should predominate has been a mystery; however our picture of the condensate vacuum state together with an effective action of the form (1) seems to provide a natural explanation for the appearance of certain non-abelian gauge symmetries. Indeed we can immediately see that phenomenological gauge symmetries may be related to the closest packing of spheres in four dimensions.

In the model vacuum of ref.2 non-linear gravitons are represented by self-dual and anti-self-dual ALE spaces. However if the Dirac strings (solotonic 7-branes in our current model) that are needed to make the metric of an ALE space well defined at infinity terminate on nearby anti-monopole singularities, then the global topology at infinity will no longer be consistent with that required for an ALE space. This suggests that the simple model of ref.2 is not completely consistent because the only self-dual or anti-self-dual solutions of the Einstein equations for which the global topology of spacetime at infinity is trivial is flat spacetime. One obvious way to resolve this conflict is to assume that the self-dual and anti-self-dual gravitons in the model of ref.2 actually correspond to compact self-dual and anti-self-dual Einstein spaces. However, the only compact self-dual or anti-self-dual solutions of the vacuum Einstein equations are the 4-torus and the K3 surface. The metric for the 4-torus contains no monopole-like singularities, and therefore is not of interest to us. The classical metric for a K3 surface is unknown, but locally a K3 surface looks like an ALE space with its non-trivial 2-cycles. As in the ALE case one could presumably remove metric singularities associated with these non-trivial 2-cycles using Dirac strings, except that now the Dirac strings originating in a self-dual K3 manifold can terminate on an anti-self-dual K3 manifold. Thus it would probably be more consistent to suppose that the actual vacuum state of quantum gravity contains a coherent superposition of pairs of self-dual and anti-self-dual compact K3 surfaces.

It is clear that when the degrees of freedom associated with the 5-form field $F_5$ are taken into account many additional degrees of freedom besides those needed for dilatonic gravity will appear. In fact our suggestion that these degrees of freedom can be identified with those of a Yang-Mills gauge theory is completely consistent with recent proposals for how gauge theories arise from F-theory. It has been shown in the case of 12-dimensional F-theory that a putative coupling of parallel 7-branes to accumulations of vanishing 2-cycles in an ALE space can yield N=1 Yang-Mills theories with A, D, or E type gauge groups [14]. Therefore it appears that the exact quantum dynamics of supersymmetric gauge theories can be encoded into the local geometry of K3 surfaces with A, D, or E type singularities. One deficiency of F-theory though is that there is no explanation why particular gauge symmetries, e.g. phenomenological gauge symmetries, might be favored. In our approach, on the other hand, it is quite plausible an action like (1) will give rise to particularly favorable geometric arrangements of 7-branes and K3 surfaces in the condensate ground state. Indeed given the important role played by 4-dimensional spaces in our description of the quantum gravity vacuum it would not be surprising if the favored gauge group was simply a reflection of the closest packing of spheres in 4-dimensions.

## 5. Conclusion

Perhaps the most remarkable immediate consequence of our approach to understanding the origin of gravity and matter is that, in contrast with generic superstring models, we can provide a natural explanation why a classical spacetime with three space and one time dimension seems to play a very special role in physics. Indeed, the order parameter for our condensate vacuum is just the classical metric for 4-dimensional spacetime! Classical K3 surfaces apparently do play an important role in describing the structure of our ground state, but only insofar as they are useful for describing the degrees of freedom in the ground state, and not as in compactified superstring models as fibers of a classical background spacetime. Thus although the action principle (1) resembles an effective action introduced to describe certain low energy interactions of superstrings compactified on a Calabi-Yau 3-fold, there is a fundamental difference between our point of view concerning the origin of this action and the point of view of superstring theory. In particular we regard (1) as an effective action for describing the collective excitations of the vacuum state and not as a mechanical action. Our view is that constructs such as quantum Yang-Mills theory are to be regarded as reflections of the correlations in our condensate ground state.

On the other hand as we have noted several times there are interesting similarities between our model for the vacuum state of quantum gravity, the theory of relativistic membranes, and F-theory. For example, the special role played by thirteen dimensions in the formulation of the action (1) suggests that this action must be related in some way to F-theory. It is interesting to note in this connection that for a fixed value of stochastic time the statistical partition function for an ensemble of interacting K3 surfaces becomes a Polyakov-like path integral for a 4-manifold propagating in a higher dimensional spacetime. Thus low lying excitations of our vacuum state do have an F-theory-like mechanical interpretation. Moreover geometric quantization of self-dual Einstein spaces yields membrane-like dynamics [3], which suggests that at least the gravitational sector of our theory has some interpretation in terms of membranes. This later question is related to the algebraic structure underlying quantization of our effective action, which is probably related to the $W_\infty$ symmetry, and undoubtedly will prove to be interesting.

The $F_4 \wedge F_5 \wedge F_4$ structure of our effective action is intriguingly reminiscent of the construction of the Golay code from three 8-bit Hamming codes [24] (or equivalently the Leech lattice from three $E_8$ root lattices [25]). It is also interesting to note in this connection that our ansatz (4) for the 4-forms $F_4$, which involved extending an SO(3) gauge connection in 3+1 dimensions to an SO(8) connection in 7+1 dimensions, has an intimate connection with octonians, which of course provides a connection with the $E_8$ root lattice. The author has previously suggested [26] that a theory unifying the various superstring theories would involve the Leech lattice, and that the Monster sporadic group would be a symmetry of this underlying theory. The main motivation for this conjecture was the observation that in the construction of the lowest dimensional non-trivial representation of the Monster group spaces of dimension 12 x $2^{12}$ play a crucial role, and these spaces can be interpreted as combining the massless states of heterotic and Type II superstrings. All of this suggests that quantization of our effective action may well reveal not only interesting algebraic structures, but also unsuspected connections between the topology of 4-manifolds and algebra.

**Acknowledgments**

The author would like to thank Bernard Grossman and Carlos Castro for valuable discussions.
**References**

1. G. Chapline, Mod. Phys. Lett. A5 (1990) 2165.
2. G. Chapline, "Quantum Model for Spacetime", Mod. Phys. Lett. A7, (1992)1959; "Anyons and Coherent States for Gravitons" in Proceedings of the XXI International Conference on Differential Geometric Methods in Theoretical Physics, edited by C. N. Yang, M. L. Ge, and X. W. Zhou (World Scientific, Singapore,1993).
3. K. Yamagishi and G. Chapline, Class. Quantum Grav. 8 (1991) 427.
4. C. Castro, Phys. Lett. B413 (1997) 53.
5. J. H. Schwarz, Phys. Lett. B360 (1995) 13; P. Horava and E. Witten, Nucl. Phys. B460 (1996) 506.
6. E. Bergshoeff, E. Sezgin, Y. Tanni, and P. K. Townsend, Ann. Phys. 199 (1990) 340.
7. E. G. Floratos and G. K. Leontaris, Phys. Lett. B223 (1989) 153.
8. C. Castro, Phys. Lett. B288 (1992) 291.
9. G. Chapline and K. Yamagishi, Phys. Rev. Lett. 66 (1991) 3064.
10. C. Castro, J. Math. Phys. 34 (1993) 681.
11. C. Castro, Chaos, Solitons, and Fractals 7 (1996) 711.
12. E. G. Floratos et. al. , "On The Instanton Solutions Of The Self-Dual Membrane In Various Dimensions", hep-th/9711044.
13. C. Vafa, Nucl. Phys. B469 (1996) 403.
14. S. Katz and C. Vafa, Nucl. Phys. B497 (1997)196.
14. A. Kumar and C. Vafa, "U-Manifolds", hep-th/9611007.
16. S. Ferrara, R. Minasian, and A. Signotti, Nucl. Phys. B474 (1996) 323.
17. E. Witten, Phys. Lett. 77B (1978) 394.
18. S. A. Merkulov, Comm. Math. Phys. 93 (1984) 325.
19. R. J. Baston and L. J. Mason, Class. Quantum Grav. 4 (1987) 815.
20. B. Grossman, T. W. Kephart, and J. Stasheff, Comm. Math. Phys. 96 (1984) 431; Phys. Lett. B220 (1989) 431.
21. A. Ashtekar, Phys. Rev. D36 (1987) 1587.
22. L. Baulieu and I. Singer, Nucl. Phys. Proc. Sup. B15 (1988) 12.
23. G. Chapline and B. Grossman, Phys. Lett. B223 (1989) 336.
24. J. H. Conway and N.J.A. Sloane, Sphere Packings, Lattices and Groups (Springer-Verlag 1988).
25. J. Lepowsky and A. Meurman, J. of Algebra 77 (1982) 484.
26. G. Chapline, Phys. Lett. B158 (1985).
12